\documentclass[11pt]{article}

\usepackage{scicite}

\usepackage{times}
\usepackage{graphicx}

\newenvironment{sciabstract}{%
\begin{quote} \bf}
{\end{quote}}

\newcounter{lastnote}

\title{Mapping Spiral Structure on the far side of the Milky Way}

\author
{Alberto Sanna,$^{1\ast}$ Mark J. Reid,$^{2}$ Thomas M. Dame,$^{2}$\\
 Karl M. Menten,$^{1}$ Andreas Brunthaler$^{1}$\\
\\
\normalsize{$^{1}$Max-Planck-Institut f\"{u}r Radioastronomie, Auf dem H\"{u}gel 69, 53121 Bonn, Germany,}\\
\normalsize{$^{2}$Harvard-Smithsonian Center for Astrophysics, 60 Garden Street, Cambridge, MA 02138, USA}\\
\\
\normalsize{$^\ast$E-mail:  asanna@mpifr-bonn.mpg.de}
}

\date{}

\begin{document}

\baselineskip24pt

\maketitle

\begin{sciabstract}
Little is known about the portion of the Milky Way lying beyond the Galactic center at distances of more
than 9\,kilo-parsec from the Sun. These regions are opaque at optical wavelengths due to absorption by
interstellar dust, and distances are very large and hard to measure. We report a direct trigonometric
parallax distance of 20.4$_{-2.2}^{+2.8}$\,kilo-parsec obtained  with the Very Long Baseline Array to a
water maser source in a region of active star formation. These measurements allow us to shed light on
Galactic spiral structure by locating the Scutum-Centaurus spiral arm as it passes through the far side of
the Milky Way, and to validate a kinematic method for determining distances in this region based on
transverse motions.
\end{sciabstract}

Large scale maps of the Milky Way have had to exclude a triangular region
with the Sun at its apex and extending $\sim$10$^{\circ}$ to either side
of the Galactic center (e.g., \cite{Oort1958}). In this region the Galaxy is
rotating almost perpendicularly to our line-of-sight, so the radial velocities
that enable the estimate of so-called kinematic distances are degenerate 
at values around zero. Some progress has been made in the nearer parts 
of this region with distances determined through stellar spectra and, more
recently, parallax measurements for astronomical sources of maser emission
from newly formed stars \cite{Reid2014}. But, the region behind the Galactic
center has remained largely out of reach due to the large distances
($>$\,9\,kpc), very high foreground extinction, and source confusion.  

The Bar and Spiral Structure Legacy (BeSSeL) Survey is a Key Project of the Very Long 
Baseline Array (VLBA), which has been measuring trigonometric parallaxes and motions
on the sky plane (hereafter, proper motions) of methanol and water masers associated
with hundreds of high-mass star forming regions in the Galaxy. Recently, the project 
focused mainly on distant regions to locate distant segments of the Galactic spiral arms.
Here, we discuss the most distant of these sources, G007.47$+$00.05 (or IRAS\,17591$-$2228),
for which an accurate parallax distance was not previously available. The source lies on the far side of
the Galaxy and appears to link the prominent Scutum-Centaurus spiral arm in the fourth
Galactic quadrant with a distant molecular arm recently discovered in the first quadrant,
dubbed the Outer Scutum-Centaurus (OSC) arm \cite{Dame2011}. 

We used the VLBA to observe strong H$_2$O maser emission at 22.2\,GHz from the 
star-forming region  G007.47$+$00.05 between 2014 March and 2015 March \cite{SupplMat}. 
We modeled the relative position in time of compact maser spots,
with respect to an extragalactic continuum source (Figure\,1A), as the sum of the parallax signature
(i.e., a sinusoid), caused by our changing vantage point as the Earth orbits the Sun, plus a linear motion
of the star-forming region relative to the Sun in the E--W and N--S directions (Figure\,1B).  
After removing the linear motion, the parallax effect inferred from this analysis (Figure\,1C) yields a
parallax angle of $0.049 \pm 0.006$\,milli-arcsecond (mas).  This measurement corresponds to 
a distance of 20.4$_{-2.2}^{+2.8}$\,kpc (66,500 ly) with an uncertainty of less than $\pm14$\%. 
In Figure\,2, we plot the Galactic position of G007.47$+$00.05, superposed on a plan view of the 
Milky Way, which shows the locations of prominent star-forming sites determined from trigonometric
parallaxes. 

We compare this result with an independent, statistical (Bayesian), distance estimate \cite{Reid2016a},
derived by taking into account complementary information about the spatial distribution and kinematics
of giant molecular clouds in the Galaxy. In Figure\,3, we plot the probability density function (PDF) for
the source distance, based on five distinct contributions (listed below), and the cumulative best estimate
for the probability of any given distance (black solid line).  The red solid line (PDF$_{\rm SA}$) quantifies
the probability that a set of three values, Galactic longitude ($\ell$), latitude ($b$), and radial velocity with
respect to the local standard of rest (LSR), V$_{\rm LSR}$, match those values expected for a given spiral
arm segment, as traced by CO and H\,{\sc i} emission (e.g. \cite{Dame2001}). The loci of individual spiral
arms are then fixed in position by trigonometric parallaxes, following previous analysis \cite{Reid2014}. Also
plotted is the kinematic distance probability (PDF$_{\rm KD}$), as inferred from the radial velocity of the
source with respect to the Sun (V$_{\rm LSR}$),  after assuming a rotation curve for the Galaxy \cite{Reid2014}. 
Because star-forming regions are expected to lie close to the Galactic plane, we have also quantified the
expectation of finding a source at high Galactic latitude if it is near the Sun, and vice versa  (PDF$_{\rm GL}$). 
Details of how these contributions are calculated have been previously discussed \cite{Reid2016a}.

In Figure\,3, we assign two additional distance PDFs to the source, based on its proper motion in the Galactic
latitude and longitude directions. The more distant the source, the smaller one expects the angular motion
out of the plane (PDF$_{\rm PM,b}$).  At the same time, the proper motion along the plane (i.e., in Galactic
longitude) provides an additional kinematic distance probability (PDF$_{\rm PM,\ell}$), which makes use of
the velocity component (V$_{\perp}$) perpendicular to the line-of-sight (e.g. \cite{Sofue2011}). The
probability density function, PDF$_{\rm PM,\ell}$, is generated similarly to the PDF$_{\rm KD}$
 \cite[their section 2.2]{Reid2016a}. The observed proper motion components ($\mu_{\rm x}$,\,$\mu_{\rm y}$)
are rotated about the line-of-sight into Galactic components ($\mu_{\ell}$,\,$\mu_{b}$), and the component
projected along the Galactic longitude ($\mu_{\ell}$) is compared with that expected under the assumption of
circular Galactic orbits as a function of distance (e.g. eq.\,4 of \cite{Sofue2011}). This calculation of kinematic
distance using proper motions is most effective toward the Galactic center and anticenter, the directions in
which conventional kinematic distances based on radial velocities are least useful. A similar calculation was previously
followed by \cite{Yamauchi2016}, who estimated for G007.47$+$00.05 a far kinematic distance of $20\pm2$\,kpc
(see supplementary online text). These authors made use of the Japanese VERA (VLBI Exploration of Radio Astrometry)
 array to observe the H$_2$O maser emission between March 2009 and December 2011.

Evaluating the kinematic distance PDFs requires Bayesian priors for the Galactic rotation curve, 
$\Theta(R)$, and the fundamental Galactic parameters of the circular rotation speed
$\Theta_0$ at the distance R$_0$ of the Sun from the Galactic center. In Table\,1,
we list the priors and source input data used to calculate the PDFs. For the V$_{\rm LSR}$
of the source, we adopt the average velocity of the H$_2$O maser emission of
$-16$\,$\pm$\,4\,km\,s$^{-1}$, which is consistent with the peak of radio recombination
lines at $-16.9$\,$\pm$\,0.1\,km\,s$^{-1}$ \cite{Balser2011} and the range of CS emission
at $-13.9$\,$\pm$\,4.6\,km\,s$^{-1}$ \cite{Bronfman1996}. For the proper motion of the star-forming 
region ($\mu_{\rm x}$,\,$\mu_{\rm y}$), we averaged the final values derived from 
Figure\,1 ($-2.42$\,$\pm$\,0.02\,mas\,yr$^{-1}$, $-4.64$\,$\pm$\,0.30\,mas\,yr$^{-1}$) 
with those measured by \cite{Yamauchi2016} based on three masers ($-2.46$\,$\pm$\,0.14\,mas\,yr$^{-1}$,
$-4.38$\,$\pm$\,0.14\,mas\,yr$^{-1}$). Since residual atmospheric delays between the target and
calibration source usually dominate the uncertainty  in proper motion measurements, and the three masers
likely have the same systematic uncertainties, we have inflated by $\sqrt{3}$ the proper motion
uncertainties from \cite{Yamauchi2016}, to account for fully correlated motions among the masers.
This analysis provides a proper motion in Galactic longitude ($\mu_{\ell}$) of $-5.12\pm0.18$\,mas\,yr$^{-1}$.

In Figure\,3, we show that the combined Bayesian distance of 20.4\,$\pm$\,0.6\,kpc agrees
well with our trigonometric parallax measurement. However, some caveats should be noted.  
Kinematic distances hold under the assumption of purely circular orbits.  This condition is 
not satisfied within about 4\,kpc of the Galactic center, where the gravitational effects
of the Milky Way's central bar cause highly non-circular motions (e.g. \cite{Binney1991,Sanna2014}). 
A line of sight from the Sun at $\ell$\,$=$\,$7.47^{\circ}$  passes through this region
between 4.7 and 12.3\,kpc; hence this range of distances cannot be excluded based on kinematic PDFs.
Our direct parallax measurement does not suffer this limitation.

The spiral arm PDF (PDF$_{\rm SA}$) has two main peaks between 4 and 5\,kpc, because the target
source has a V$_{\rm LSR}$ between that of the Norma ($\sim$10\,km\,s$^{-1}$) and Near 3-kpc 
($-20$\,km\,s$^{-1}$) arms at that longitude (e.g. Figure\,7 of \cite{Sanna2014});
the current parallax measurement unambiguously rules out such small distances. 
Distances to spiral arms beyond about 10\,kpc are a substantial extrapolation because 
of the current lack of direct parallax measurements. The spiral arm PDF shows a
small peak at $\sim$16 kpc based on a long extrapolation of the Outer arm (red curve
in Figure\,2) through the first Galactic quadrant. However, locating G007.47$+$00.05 at
this distance can be excluded by the parallax distance, with a probability of  $>95$\%,  as well
as by kinematic distance values, since this portion of the Outer arm is well outside the Galactic
center region.    

Close to the Galactic plane at a longitude of $7.5^{\circ}$ (e.g. Figure\,3 of \cite{Dame2011}),  
the OSC arm has a V$_{\rm LSR}$ ($-12$\,km\,s$^{-1}$) close to that of the target source. 
If the Scutum-Centaurus arm is a logarithmic spiral with constant pitch angle ($\psi$) of
$14^{\circ}$, which is constrained by the tangent directions of the Scutum ($\ell$\,$=$\,$31^{\circ}$)
and Centaurus ($\ell$\,$=$\,$309^{\circ}$) arms in addition to nearby parallax measurements
\cite{Dame2011}, the distance to the OSC arm at the longitude of the maser source would be in
the range 20--21\,kpc. In Figure\,2, we show the logarithmic spiral of the OSC arm under these
conditions. This arm also coincides, within the uncertainties, with the farthest arm traced by 
\cite{Nakanishi2016}, taking into account the value of R$_0$ adopted by those authors. This distance
estimate determines a third peak of the PDF$_{\rm SA}$  which is consistent with the trigonometric
parallax distance. In Figure\,3, the combination of spiral and (both) kinematic PDFs strongly constrains
a narrow range of distances (absolute peak of the black solid line), which matches the trigonometric
distance of 20.4\,kpc. 

This result has two further consequences.  First, the agreement of the parallax and 
the kinematic distances can be interpreted as evidence that the approximation of 
circular motions holds at Galactocentric radii near 12\,kpc in the far outer Galaxy. 
Second, it calls into question the assumption of constant pitch angle of the Outer
Scutum-Centaurus arm along its winding around the Galaxy.  In Figure\,2, we extend the track of the 
Scutum segment from the first Galactic quadrant through the fourth quadrant, by assuming 
a logarithmic spiral with $\psi$ of $20^{\circ}$. This pitch angle provides the best 
spiral fit to trigonometric distances of star-forming regions in the Scutum arm between 
$\ell$ of $5.9^{\circ}$ and $32.0^{\circ}$ \cite{Sato2014,Krishnan2015}. Because the near 
segment of the Scutum-Centaurus arm (i.e., the Scutum arm in the first quadrant) has
a much larger pitch angle than that determined from the Scutum and Centaurus tangents
and the location of G007.47+00.05 ($14^{\circ}$), it is clear that a single logarithmic spiral
cannot describe the full complexity of the Scutum-Centaurus-OSC arm. This result
is in agreement with a recent analysis of the spiral arm morphology in four external 
face-on galaxies, where pitch angles can vary along individual arms by more than 
$10^{\circ}$ \cite{Honig2015}.

In summary, we have measured the distance to a water maser source in the OSC arm using both a
direct trigonometric parallax and a statistical Bayesian analysis. Our parallax distance agrees with
an indirect distance estimate for the source inferred from its measured (proper) motion on the sky
\cite{Yamauchi2016}, thus providing support for the determination of kinematic distances based
on proper motions rather than radial motions alone. These measurements allow us to trace the 
Scutum-Centaurus arm nearly one full turn of Galactic azimuth and out to large Galactocentric radii
on the far side of the Milky Way, and shows that we can map spiral structure throughout the 
Galaxy.   

\clearpage


\bibliography{scibib}

\begin{thebibliography}{10}

\bibitem{Oort1958}
J.~H. {Oort}, F.~J. {Kerr}, G.~{Westerhout}, {\it Mon. Not. R. Astron. Soc.\/}
  {\bf 118}, 379 (1958).

\bibitem{Reid2014}
M.~J. {Reid}, {\it et~al.\/}, {\it Astrophys. J.\/} {\bf 783}, 130 (2014).

\bibitem{Dame2011}
T.~M. {Dame}, P.~{Thaddeus}, {\it Astrophys. J.\/} {\bf 734}, L24 (2011).

\bibitem{SupplMat}
Materials and methods are available as supplementary materials on Science Online.

\bibitem{Reid2016a}
M.~J. {Reid}, T.~M. {Dame}, K.~M. {Menten}, A.~{Brunthaler}, {\it Astrophys.
  J.\/} {\bf 823}, 77 (2016).

\bibitem{Dame2001}
T.~M. {Dame}, D.~{Hartmann}, P.~{Thaddeus}, {\it Astrophys. J.\/} {\bf 547},
  792 (2001).

\bibitem{Sofue2011}
Y.~{Sofue}, {\it Publ. Astron. Soc. Jpn.\/} {\bf 63}, 813 (2011).

\bibitem{Yamauchi2016}
A.~{Yamauchi}, {\it et~al.\/}, {\it Publ. Astron. Soc. Jpn.\/} {\bf 68}, 60
  (2016).

\bibitem{Balser2011}
D.~S. {Balser}, R.~T. {Rood}, T.~M. {Bania}, L.~D. {Anderson}, {\it Astrophys.
  J.\/} {\bf 738}, 27 (2011).

\bibitem{Bronfman1996}
L.~{Bronfman}, L.-A. {Nyman}, J.~{May}, {\it Astron. Astrophys. Suppl. Ser.\/}
  {\bf 115}, 81 (1996).

\bibitem{Binney1991}
J.~{Binney}, O.~E. {Gerhard}, A.~A. {Stark}, J.~{Bally}, K.~I. {Uchida}, {\it
  Mon. Not. R. Astron. Soc.\/} {\bf 252}, 210 (1991).

\bibitem{Sanna2014}
A.~{Sanna}, {\it et~al.\/}, {\it Astrophys. J.\/} {\bf 781}, 108 (2014).

\bibitem{Nakanishi2016}
H.~{Nakanishi}, Y.~{Sofue}, {\it Publ. Astron. Soc. Jpn.\/} {\bf 68}, 5 (2016).

\bibitem{Sato2014}
M.~{Sato}, {\it et~al.\/}, {\it Astrophys. J.\/} {\bf 793}, 72 (2014).

\bibitem{Krishnan2015}
V.~{Krishnan}, {\it et~al.\/}, {\it Astrophys. J.\/} {\bf 805}, 129 (2015).

\bibitem{Honig2015}
Z.~N. {Honig}, M.~J. {Reid}, {\it Astrophys. J.\/} {\bf 800}, 53 (2015).



\bibitem{Elitzur1992}
M.~{Elitzur}, ed., {\it {Astronomical masers}\/}, vol. 170 of {\it Astrophysics
  and Space Science Library\/} (1992).

\bibitem{ReidHonma2014}
M.~J. {Reid}, M.~{Honma}, {\it Annu. Rev. Astron. Astrophys.\/} {\bf 52}, 339
  (2014).

\bibitem{Deller2007}
A.~T. {Deller}, S.~J. {Tingay}, M.~{Bailes}, C.~{West}, {\it Publ. Astron. Soc.
  Pac.\/} {\bf 119}, 318 (2007).

\bibitem{vanMoorsel1996}
G.~{van Moorsel}, A.~{Kemball}, E.~{Greisen}, {\it Astronomical Data Analysis
  Software and Systems V\/}, G.~H. {Jacoby}, J.~{Barnes}, eds. (1996), vol. 101
  of {\it Astronomical Society of the Pacific Conference Series\/}, p.~37.

\bibitem{Reid2009a}
M.~J. {Reid}, {\it et~al.\/}, {\it Astrophys. J.\/} {\bf 693}, 397 (2009).

\bibitem{Kettenis2006}
M.~{Kettenis}, H.~J. {van Langevelde}, C.~{Reynolds}, B.~{Cotton}, {\it
  Astronomical Data Analysis Software and Systems XV\/}, C.~{Gabriel},
  C.~{Arviset}, D.~{Ponz}, S.~{Enrique}, eds. (2006), vol. 351 of {\it
  Astronomical Society of the Pacific Conference Series\/}, p. 497.

\bibitem{Honma2008}
M.~{Honma}, Y.~{Tamura}, M.~J. {Reid}, {\it Publ. Astron. Soc. Jpn.\/} {\bf
  60}, 951 (2008).

\bibitem{Wu2014}
Y.~W. {Wu}, {\it et~al.\/}, {\it Astron. Astrophys.\/} {\bf 566}, A17 (2014).

\bibitem{Sanna2012}
A.~{Sanna}, {\it et~al.\/}, {\it Astrophys. J.\/} {\bf 745}, 191 (2012).

\bibitem{Trinidad2013}
M.~A. {Trinidad}, {\it et~al.\/}, {\it Mon. Not. R. Astron. Soc.\/} {\bf 430},
  1309 (2013).

\bibitem{Burns2016}
R.~A. {Burns}, T.~{Handa}, T.~{Nagayama}, K.~{Sunada}, T.~{Omodaka}, {\it Mon.
  Not. R. Astron. Soc.\/} {\bf 460}, 283 (2016).

\bibitem{Sofue1999}
Y.~{Sofue}, {\it et~al.\/}, {\it Astrophys. J.\/} {\bf 523}, 136 (1999).

\bibitem{Fey2004}
A.~L. {Fey}, {\it et~al.\/}, {\it Astron. J.\/} {\bf 127}, 3587 (2004).

\bibitem{Immer2011}
K.~{Immer}, {\it et~al.\/}, {\it Astrophys. J. Suppl. Ser.\/} {\bf 194}, 25
  (2011).

\bibitem{Sanna2010}
A.~{Sanna}, {\it et~al.\/}, {\it Astron. Astrophys.\/} {\bf 517}, A71 (2010).



\end{thebibliography}

\bibliographystyle{Science}



\noindent \textbf{Acknowledgements:} A.S. gratefully acknowledges financial support by the
Deutsche Forschungsgemeinschaft (DFG) Priority Program 1573. The National Radio Astronomy
Observatory is a facility of the National Science Foundation operated under cooperative agreement
by Associated Universities, Inc. This work made use of the Swinburne University of Technology
software correlator, developed as part of the Australian Major National Research Facilities
Programme and operated under license. The authors thank M.\,Honma
for fruitful discussions in preparation.
\textbf{Data and materials availability:} All data used in the paper are public available through the
National Radio Astronomy Observatory (NRAO) archive under program BR198, at 
\emph{$https://archive.nrao.edu/archive/advquery.jsp$}. A spectrum of the maser emission on
2014 March 18 can be found at \emph{$http://bessel.vlbi-astrometry.org/first\_epoch$}.

\noindent \textbf{Supplementary Materials} \\
www.sciencemag.org \\
Materials and Methods \\
Supplementary Text \\
Figures 4, 5 \\
Tables 2, 3, 4 \\
References (17--31) [Note: The numbers refer to any additional references cited only within the Supplementary Materials]

\clearpage


\begin{figure*}
\centering
\includegraphics[angle=0,scale=.6]{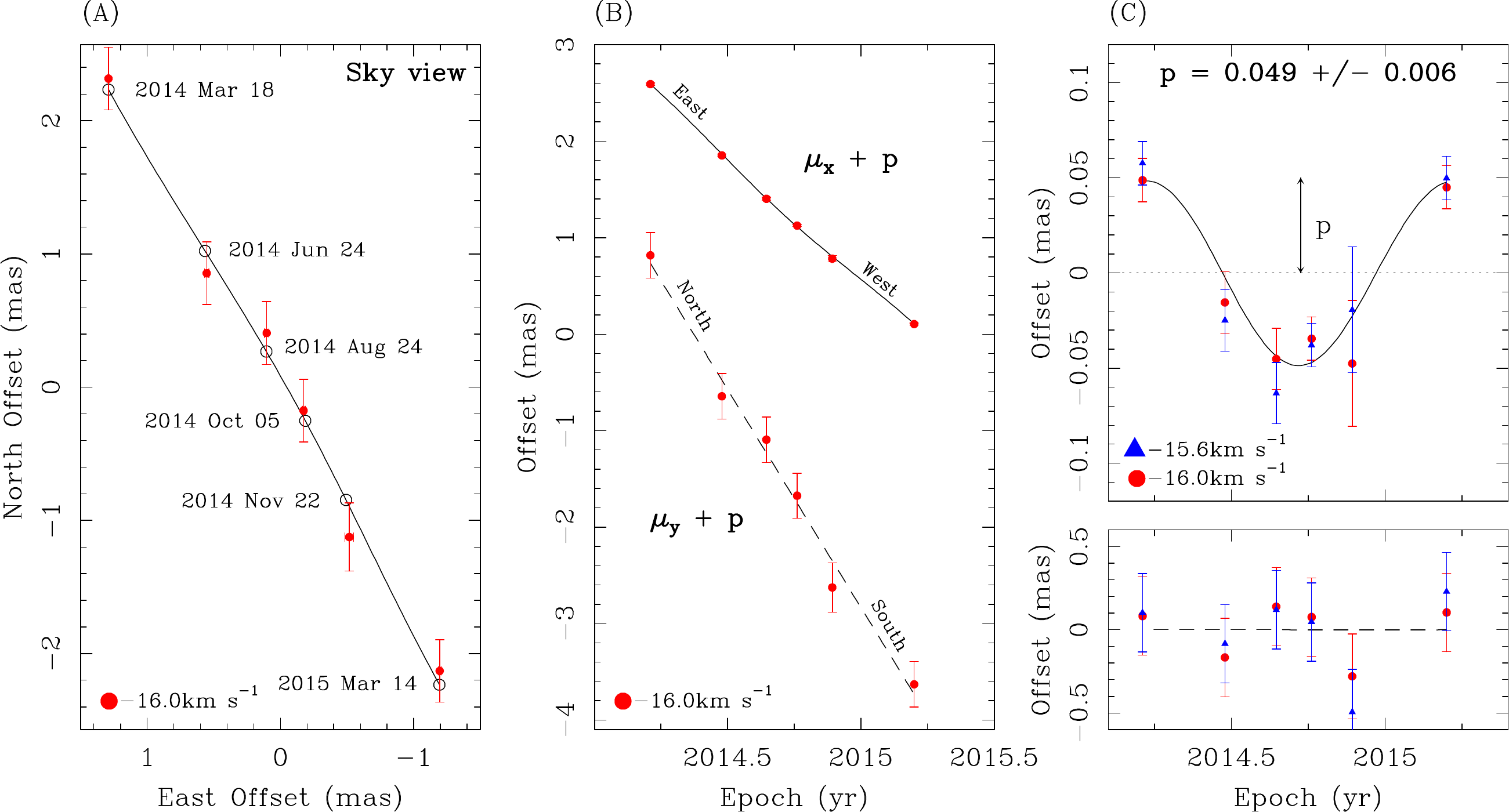}
\caption{\textbf{Results of the combined parallax fit for G007.47$+$00.05.} This fit is obtained from the combination of two
point-like maser spots with respect to the background quasar J175526.28$-$223210.6 (or [IBR2011]\,J1755$-$2232; Table\,S2). For clarity, we draw only maser
positions for the spot at  V$_{\rm LSR}$ of --16.0\,km\,s$^{-1}$ (red circles). (\textbf{A}) Sky projected motion of the
maser source with respect to J175526.28$-$223210.6 with each epoch labeled (the zero position is arbitrary). The empty circles and the
line show the best-fitting position offsets and the trajectory, respectively. East offset uncertainties are smaller than the symbol.
(\textbf{B}) Decomposed offset positions for the maser source along the east and north directions versus time. The best-fitting
models in east and north directions are shown as continuous and dashed lines, respectively.  (\textbf{C}) Same as the middle
panel but with fitted proper motions  subtracted ($\mu_{\rm x}$ and $\mu_{\rm y}$), yielding the parallax sinusoid. Positions
for the spot at V$_{\rm LSR}$ of --15.6\,km\,s$^{-1}$ are also shown for the final parallax (p).  The east offset (upper panel)
and north offset (lower panel) data are shown separately on different scales. \label{parallax}}
\end{figure*}


\begin{figure}
\centering
\includegraphics[angle=0,scale=.5]{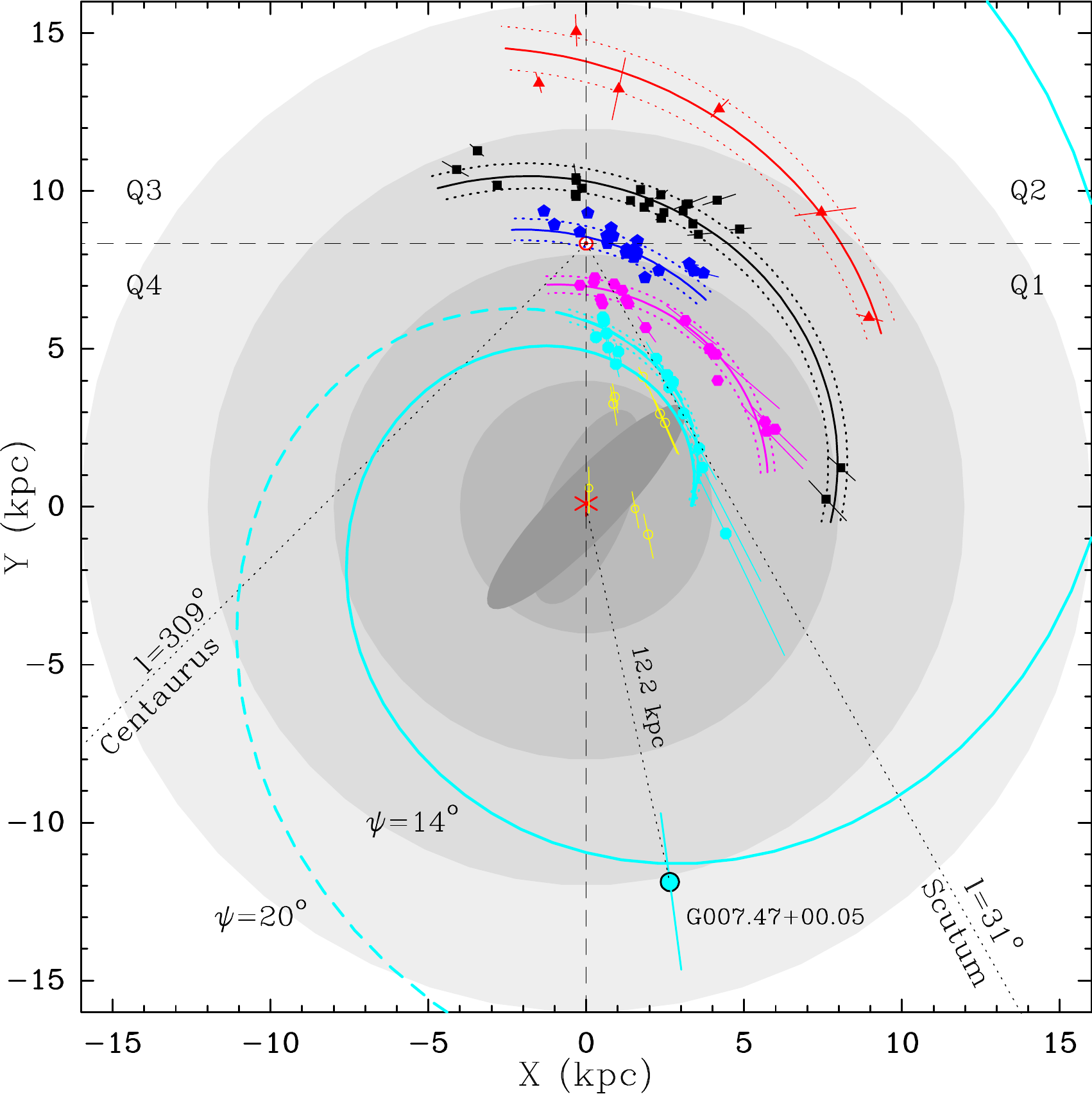}
\caption{\textbf{Plan view of the Milky Way showing the location of G007.47$+$00.05 and other maser sources determined via trigonometric
parallaxes \cite[their Table\,1]{Reid2014}.} The reference system gives offset positions from the Galactic center (red
asterisk). Colored symbols designate distinct regions in the Galaxy: inner Galaxy sources (empty yellow circles); Scutum-Centaurus arm (solid
cyan circles); Sagittarius arm (magenta hexagons); Local arm (blue pentagons); Perseus arm (black squares); Outer arm (red triangles). Error
bars correspond to distance uncertainties of 1\,$\sigma$. The background gray disks, centered on the Galactic center, show regions within
Galactocentric radii of 4, 8, 12, and 16\,kpc. The Sun is located at (0,\,8.34)\,kpc. Galactic quadrants, centered on the Sun, are indicated
with dashed lines. Solid lines trace log-periodic spiral fits to sources grouped by arms; dotted lines correspond to 1\,$\sigma$ widths. Two
logarithmic spirals are drawn for the Scutum-Centaurus arm, one with a pitch angle ($\psi$) of $14^{\circ}$ (solid), constrained by the
Scutum and Centaurus tangent directions \cite{Dame2011}, and the other with a pitch of $20^{\circ}$ (dashed), based on maser sources
in the inner first quadrant \cite{Sato2014}. The Scutum and Centaurus tangent directions are also shown. (Adapted from \cite{Reid2014},
their Figure\,1).  
\label{galaxy}}
\end{figure}


\begin{figure*}
\centering
\includegraphics[angle=0,scale=.60]{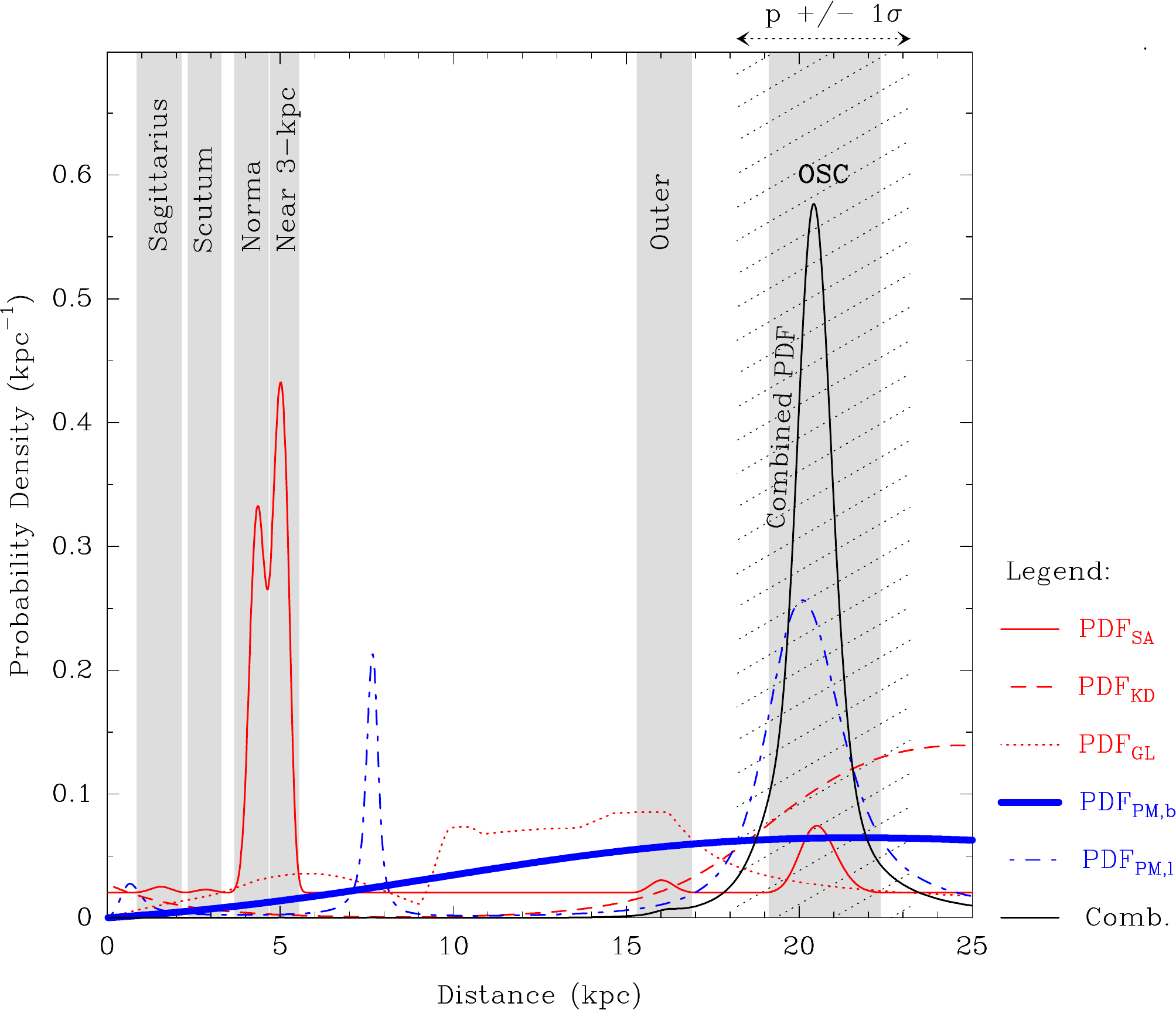}
\caption{\textbf{Distance probability density function (PDF) for the star-forming region G007.47$+$00.05.} Individual PDFs are
indicated by subscripts and refer to the spiral arm assignment of the source (SA), its kinematic distance (KD), Galactic latitude (GL),
and its proper motion in latitude (PM,\,$b$) and longitude (PM,\,$\ell$). Input parameters for the PDFs are given in Table\,1.
Grey areas mark the expected regions for each spiral arm at a Galactic longitude of $7.5^{\circ}$. The combined PDF (black solid line)
strongly favors an association with the Outer Scutum-Centaurus (OSC) arm at a distance of 20.4\,$\pm$\,0.6\,kpc. This estimate
agrees with the trigonometric parallax distance of 20.4$_{-2.2}^{+2.8}$\,kpc (hatched black area).\label{pdf}}
\end{figure*}

\clearpage


\begin{center}
\begin{table*}
{\footnotesize
\caption{\textbf{Parameters used to calculate the distance PDFs.} Column\,1 gives the type of 
Galactic rotation curve adopted for the calculation \cite[their Table\,5]{Reid2014}. Columns\,2 
and~3 report the fundamental Galactic parameters of circular rotation speed at the Solar circle and
the distance of the Sun from the Galactic center, respectively \cite{Reid2014}. Columns\,4 and~5
give the Galactic coordinates of the source in the longitude and latitude directions, respectively.
Columns\,6, 7, and~8 give the local standard of rest velocity of the source, and its motions
on the sky plane, respectively.}
\begin{tabular}{ccc|ccccc}
\hline \hline
\multicolumn{3}{c}{Priors} & \multicolumn{5}{c}{Input Data} \\
  $\Theta(R)$ &  $\Theta_0$    &  R$_0$ &      $\ell$      &       $b$        &  V$_{\rm LSR}$  &   $\mu_{x}$    &     $\mu_{y}$      \\
               & (km\,s$^{-1}$)  &  (kpc)   &  ($^{\circ}$) & ($^{\circ}$) &   (km\,s$^{-1}$) &  (mas yr$^{-1}$) &   (mas yr$^{-1}$)  \\
\hline 
Univ. & $241\pm8$ & $8.34\pm0.16$  & 7.471  & 0.058  & $-16\pm4$ & $-2.44\pm0.07$ & $-4.51\pm0.17$  \\
\hline
\end{tabular}
}
\end{table*}
\end{center}


\section*{Supplementary Materials} 

\section*{Materials and methods}

\subsection*{Observations and Calibration} \label{obs}

We conducted VLBA observations of the $6_{16}$\,$-$\,$5_{23}$ H$_2$O maser emission
line (e.g. \emph{(17)}; rest frequency 22.235080 GHz) towards the star-forming region
G007.47$+$00.05. Six separate observations were performed over the course of one year; 
observing dates are indicated in Figure\,1A. Relative astrometric accuracy is inversely proportional
to the angular separations of sources on the sky, so we selected three background sources (Table\,S1)
located $<$\,4$^{\circ}$ from the maser as calibration points. At each epoch we observed for
7 hours and calibrated the atmospheric propagation path-delays for each antenna. 
Details of the observational strategy can be found in \emph{(18)}. 
Table\,S1 summarizes the source information.

Four adjacent intermediate frequency bands, each 16\,MHz wide, were recorded in dual circular
polarization; each band was correlated to produce 32 spectral channels. The third band was centered
at the LSR velocity of the strongest maser feature ($-16.0$\,km\,s$^{-1}$).  This band was additionally
correlated with 2000 spectral channels, each 8.0\,kHz wide, in order to sample the maser linewidths
with a velocity resolution of 0.11\,km\,s$^{-1}$. The data were processed with the VLBA DiFX software
correlator \emph{(19)} using an integration time of 1\,s.

Data were calibrated and imaged with the NRAO Astronomical Image Processing System (AIPS; e.g.
\emph{(20)}) following the procedure described in \emph{(21)} and using the
ParselTongue scripting interface \emph{(22)}. For the final imaging, we removed any data for
which an antenna in an interferometer baseline had an elevation less than $15^{\circ}$ above the horizon. 
In Figure\,S1, we show the images of the background sources from the first epoch, 
phase-referenced to the maser channel at V$_{\rm LSR}$ of $-15.6$\,km\,s$^{-1}$. 

\subsection*{Parallax fitting} \label{obs}

Maser spots for parallax fitting were selected according to the following criteria: (1) spots
persisting over the 6 epochs that belong to isolated cloudlets, in order to avoid emission
blended between different maser centers; (2) compact maser spots, unresolved by the VLBA
beam or slightly resolved but with a stable, deconvolved, position angle; (3) bright maser
spots with signal-to-noise ratios of more than a hundred.

As described in the main text, we fitted the observations with an astrometric model that
accounts for both parallax and proper motion of the source. Results of the parallax and
proper motion fitting for two point-like maser spots with respect to J175526.28$-$223210.6, the
calibration source projected nearest the maser, are listed in Table\,S2. The formal fitting uncertainties
were combined in quadrature with error floors to account for the effects of unmodeled
atmospheric delays.  These error floors were set to $\pm$0.005\,mas and $\pm$0.23\,mas
in the E--W and N--S  directions, respectively, so as to yield values of reduced $\chi^2$ 
(per degree of freedom) near unity for each coordinate (e.g. \emph{(21)} for details).  
On 2014 November 22 only four antennas out of nine could be used for imaging (because of
instrumental issues), with a maximum baseline length of $\approx2000$\,km, and the formal
position uncertainties were 3 times greater than at the other epochs (Figure\,1C). We simultaneously
fitted the positions of the two spots listed in Table\,S2, solving for a single parallax, but allowing
for different proper motions. The final parallax uncertainty was multiplied by $\sqrt{2}$ to account
for the possibility of residual atmospheric delays between the maser and the calibrator line-of-sights, 
which would be fully correlated for the two masers spots.  The combined parallax
fit yielded a parallax of $0.049 \pm 0.006$\,mas and is shown in Figure\,1C.

Note that, especially for sources at low declination, N--S components of relative
positions often have greater uncertainties than the E--W components, mainly because
systematic errors from unmodeled atmospheric delays are more strongly correlated
with N--S positions (e.g. \emph{(23)}). This effect is also well established 
by previous observations (e.g. \emph{(14, 24)}), and the observing
epochs are selected to optimally sample the sinusoidal parallax signature in right
ascension.

Since the parallax uncertainty, established by the magnitude of the post-fit residuals, 
is fairly small, we now evaluate whether this is of the expected magnitude. 
At a frequency of 22\,GHz, the dominant error in relative position measurements is due to 
residual tropospheric delays, which predominantly scale with the zenith angle between sources. 
The calibrator J175526.28$-$223210.6 has a negligible N--S offset with respect to the target source 
($\theta_{\rm N}$\,$\sim$\,0, or a position angle, P.A., of $90^{\circ}$).  As shown by
\emph{(23)}, an E--W alignment between calibrator and target source minimizes
astrometric errors for low declination sources.  According to their analysis, observations 
at a declination of $-30^{\circ}$, with the four stations of the Japanese VERA
array, would provide single-epoch position errors 
of $\pm$\,0.042\,mas in the E-W direction, for an E--W separation of $1^{\circ}$ 
and 1\,cm rms vertical path-delay uncertainty.   For a source declination of
$+15^{\circ}$, astrometric errors would be reduced to $\pm$\,0.025\,mas under the same 
conditions.  At the Declination of about $-22^{\circ}$ of our sources
and a maser-calibrator separation of $1.6^{\circ}$, one would therefore expect 
relative position errors of  $\approx0.060$\,mas.  However, the VLBA provides
twice the maximum baseline length of VERA (2300\,km), as well as double the number 
of antennas, which should reduce position uncertainties by a factor $1/(2\times\sqrt{2})$. 
Therefore, for our observations, we expect an E--W parallax uncertainty of
0.060\,mas\,$\times$\,$(1/(2\times\sqrt{2}))$\,$\times$\,0.5\,$\times$\,$(1/\sqrt{6-3})$, 
where the 0.5 factor comes from the fact that we measure the full parallax sinusoid 
(i.e., double the parallax angle), and $\sqrt{6-3}$ is the degrees of freedom in the model. 
This estimate of the parallax uncertainty ($\pm$\,0.006\,mas) is in agreement with the 
measured value. 

We discarded parallax fits for the second and third calibration sources, both of which are 
projected more than 3$^{\circ}$ away from the target maser ($\theta_{\rm N}$\,$>$\,$2^{\circ}$),
because of high residual uncertainties in the E--W offsets.  If we repeat the above calculation of
expected parallax uncertainty  for the calibrator J180740.68$-$250625.9, taking into account
its offset and position angle with respect to the maser source, we obtain an E--W position error of 
$\pm$\,0.24\,mas\,$\times$\,$(1/(2\times\sqrt{2}))$ for a single epoch of VLBA observations. 
The same calculation for the calibrator J175141.34$-$195047.5 yields an E--W error of the same
order. These predicted single-epoch errors are smaller, by a factor of about 2.5,
than the error floors obtained from the parallax fit of the two calibration sources ($\geq$\,0.2\,mas). 
A posteriori, this result suggests that the astrometric errors introduced by the atmosphere 
at the large offsets of those two calibrators may be non-linear with the distance from the target.

\subsection*{Maser source and internal proper motions}\label{ppm}

In Figure\,S2, we show the distribution of individual maser emission centers (or cloudlets) detected
towards G007.47$+$00.05. Maser properties are reported in Table\,S3. 
The maser emission clusters at two positions, labeled as the northwestern (NW) and southeastern (SE)
groups, which have an average separation of about 60\,mas (or 1224\,au, at the distance of 20.4\,kpc). 
Their spatial distribution resembles the bipolar shocks previously reported in other sources \emph{(25, 26, 27)}. 
We measured the internal maser proper motions by applying a simple model of expansion 
between the NW and SE clusters of maser emission.  We calculated the centroid position of 
the two groups of masers from those cloudlets which persist over 1\,yr, and determined 
their median position at each epoch.  We then assumed this median position as the best 
guess for the star position (Figure\,S2) and measured the relative motion of the maser cloudlets with
respect to the star. These relative proper motions are drawn in Figure\,S2. The proper motion along the
east-west and north-south directions (V$_x$,\,V$_y$) of the maser spots at a V$_{\rm LSR}$ of
$-15.6$\,km\,s$^{-1}$ and $-16.0$\,km\,s$^{-1}$, which were used for the parallax measurements,
are ($-3.5$,\,$9.8$)\,km\,s$^{-1}$ and  ($-13.5$,\,$-6.2$)\,km\,s$^{-1}$,
respectively. The uncertainty on each velocity component is 0.8\,km\,s$^{-1}$. The proper motion of the
central star that excites the maser spots was derived by correcting the combined fit values for the peculiar
motion of the maser spots with respect to the central star (Table\,S2).

\section*{Supplementary text}

In the following, we comment on the proper motion measurements presented in Table\,S2
with respect to the values reported by \emph{(8, their Table\,3)}. These authors observed 
the H$_2$O maser emission from the star-forming region G007.47$+$00.05 over 3\,yr.
They conducted astrometric observations with the Japanese VERA array and measured 
the relative position in time of the H$_2$O maser source with respect to the background
calibrator J175526.28$-$223210.6, the same calibrator we observed with the VLBA. 
Because the VERA observations covered a time baseline three time longer than that of 
the VLBA observations, this difference would provide a higher accuracy in the proper 
motion measurements by a factor $\sqrt{3}$, under the same conditions. In the N--S
direction, the VERA and VLBA observations have comparable positional uncertainties 
(cf. Figure\,1C and Figure\,4 of \emph{(8)}), which are dominated  by unmodeled
atmospheric delays. Therefore, the N--S component of the proper motion ($\mu_{\rm y}$)
determined with the VERA observations is more accurate by a factor  $\sim$\,$\sqrt{3}$
than that determined with the VLBA observations. On the other hand, in the E--W 
direction, the VLBA observations are more accurate by a factor $\sim$\,20 than the
VERA observations (cf. Figure\,1C and Figure\,4 of \emph{(8)}). Taking into account the
different time baseline of the observations, the accuracy of the VLBA and VERA proper
motions eventually differs by an order of magnitude in the E--W direction.  
 
In this work, we corrected the proper motions determined from the parallax fitting 
by the internal motions of the maser distribution (see above). In the analysis by \emph{(8)},
they detected three maser cloudlets with negligible relative motions, and made a
weighted averaged of the proper motions determined from the three masers (their Table\,3).
Because of the different methods, we used an arithmetic average to combined the proper
motion determined in Table\,S2 with those by \emph{(8)}. 

The distance estimated from the proper motion measurements by \emph{(8)} is in 
agreement with the trigonometric parallax distance determined in this work. The
uncertainty estimated by \emph{(8)} for the ``far'' kinematic distance is also similar
to that derived from the trigonometric parallax. These authors explored the effects 
on the kinematic distance of changing $\Theta(R)$ and/or the Solar motion values, and  
showed that the uncertainty on the circular rotation speed at the Sun ($\Theta_0$) 
dominates the uncertainty of the kinematic distance. For example, an uncertainty of
$\pm$\,14\,km\,s$^{-1}$ on $\Theta_0$ corresponds to a distance uncertainty of
$\pm$\,1.2\,kpc. The same analysis applies to the sensitivity of the kinematic 
PDF$_{\rm PM, \ell}$ on the choice of the priors. 

On the other hand, a ``near'' kinematic distance could not be excluded a priori by \emph{(8)}.
In addition to the expectation of large non-circular velocities in the Galactic center region, 
which might be able to produce any combination of proper motion and radial velocity,
a ray from the Sun towards a longitude of 7.47$^{\circ}$ comes within about 1\,kpc 
of the Galactic center. In external galaxies similar to the Milky Way, rotation curves 
significantly drop inward at small Galactocentric radii (e.g. \emph{(28)}). Therefore, the
circular rotation speed in the Galactic center region is likely far less than the value assumed
by  \emph{(8)}.


\begin{figure*}[b]
\centering
\includegraphics[angle=0,scale=.5]{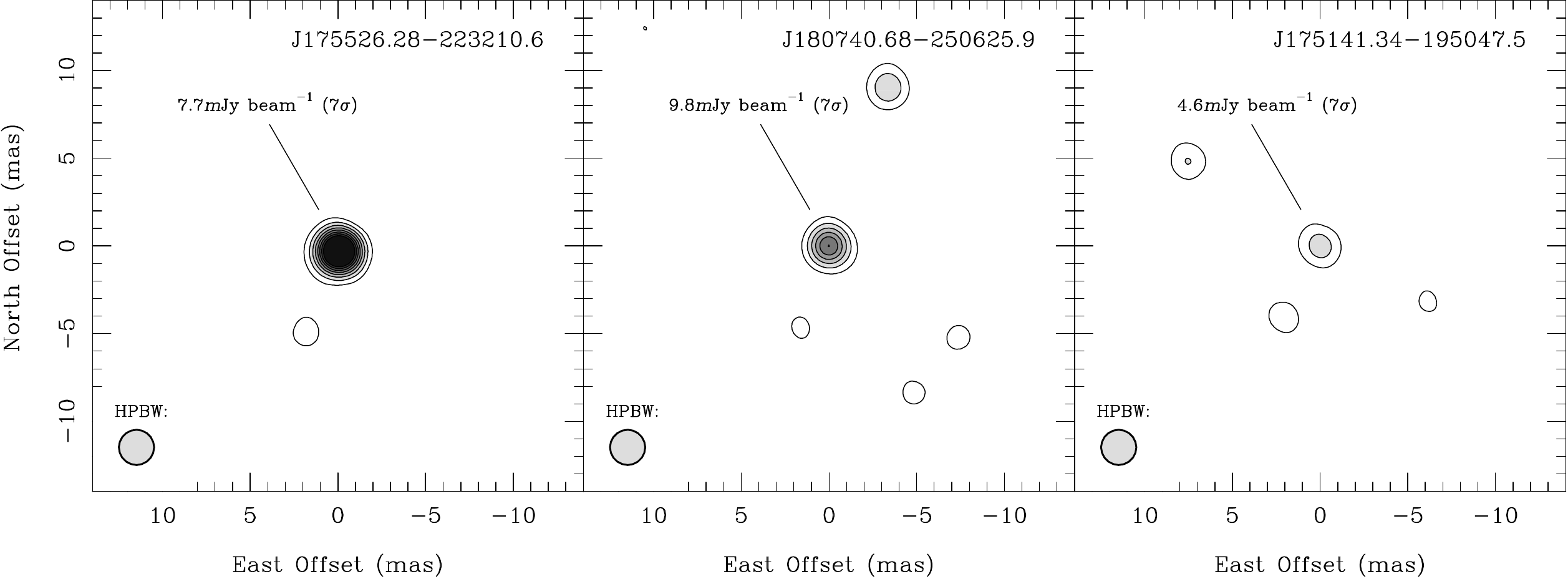}
\caption{\textbf{Images of the calibration continuum sources at 22\,GHz near the target maser.} 
The calibrators were phase-referenced to the maser channel at V$_{\rm LSR}$ of $-15.6$\,km\,s$^{-1}$.
Offsets are given with respect to the brightness peak of each calibrator. Source names are indicated in
the upper right corner and restoring beams are drawn in the lower left corner of each panel. Images are
from the first epoch observations. Contour levels start at 7$\sigma$ in steps of 7$\sigma$ (see
Table\,S1). \label{qso}}
\end{figure*}


\begin{figure}
\centering
\includegraphics[angle=0,scale=0.6]{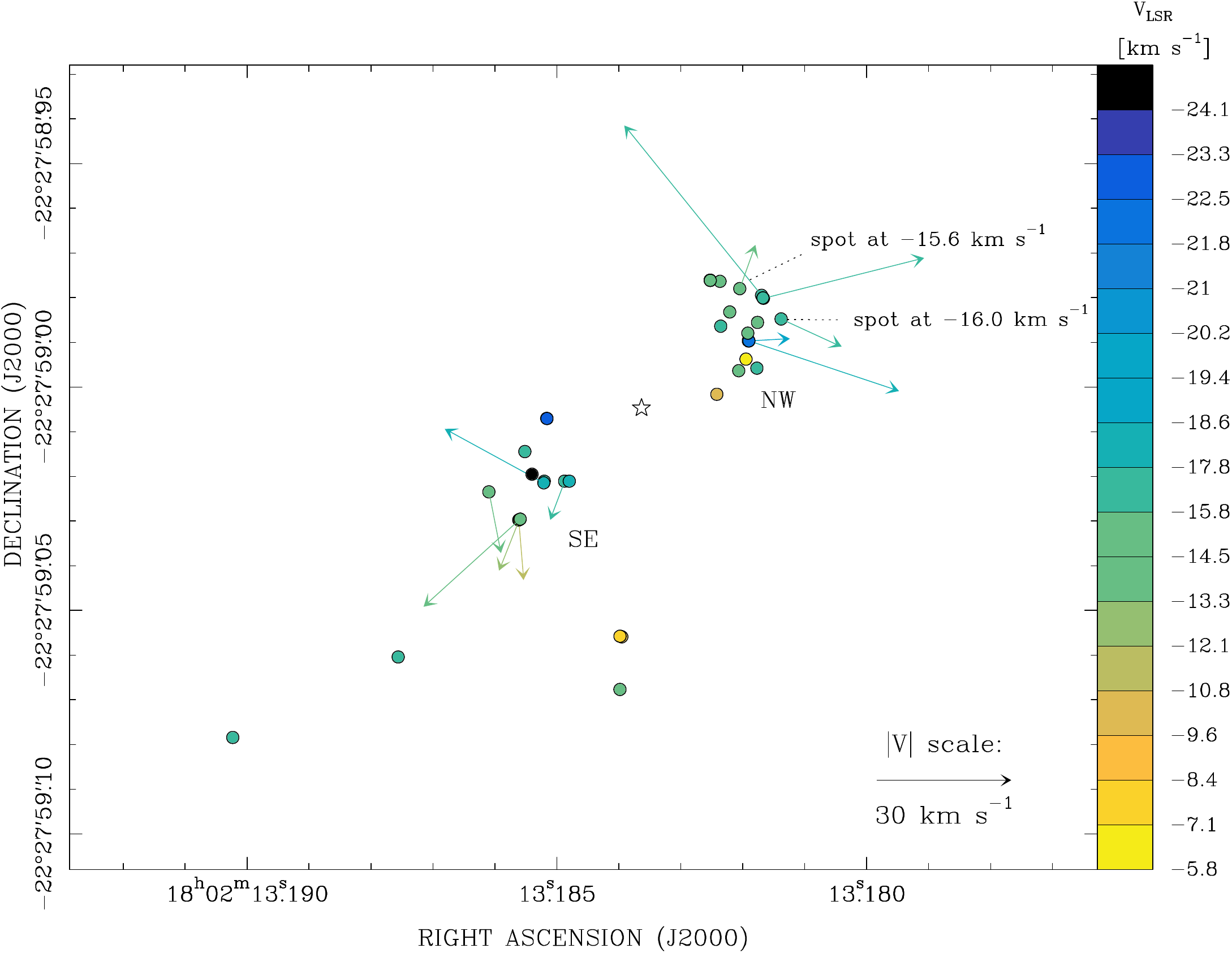}
\caption{\textbf{Distribution (dots) and proper motions (arrows) of the H$_2$O maser cloudlets detected towards G007.47$+$00.05.} The 
color of each spot indicates the gas LSR velocity as given by the color bar. The proper motion scale is shown in the bottom right corner.
The two main clusters of maser emission are labeled NW and SE. The white star between the NW and SE clusters marks the modeled
position of the central exciting star. Positions of the two spots used for the parallax fitting are indicated.
\label{internal}}
\end{figure}

\clearpage


\begin{table*}
{\footnotesize
\caption{\textbf{Source Information.} Positions and source properties for the target maser and the background calibrators
\emph{(29, 30)}, listed by increasing angular  separation. The peak position of the phase-reference maser channel
No.\,997 on 2014 March 18 (first epoch) is given in columns~2 and~3 and is accurate to within $\pm 1$\,mas. We report
the positions of the calibrators used at the VLBA correlator. Angular offsets in the east ($\theta_{\rm E}$) and north
($\theta_{\rm N}$) directions relative to the maser source are indicated in columns 4 and 5. Column~6 gives the restoring
beam size (HPBW, round) used at each epoch. Columns~7 and~8 give the peak intensities (F$_{\rm peak}$) and the image
rms noise of the phase-reference maser channel (at V$_{\rm LSR}$ of $-15.6$) and 22\,GHz background sources (from the first epoch data).}
\begin{center}
\begin{tabular}{lccrrccc}
\hline \hline
\multicolumn{1}{c}{Sources (identifier)} & R.A.\,(J2000) & Dec.\,(J2000) & $\theta_{\rm E}$ & $\theta_{\rm N}$ & HPBW & F$_{\rm peak}$ &
Image rms  \\
  & (h m s) & ($^{\circ}$ ' '') & ($^{\circ}$) & ($^{\circ}$) & (mas) & (Jy beam$^{-1}$) & (Jy beam$^{-1}$)  \\
\hline 
\scriptsize G007.47$+$00.05            (IRAS\,17591$-$2228) & 18:02:13.1820 & $-$22:27:58.978  & ...         & ...            & 2.0 & 12.67 & 0.047    \\
\scriptsize J175526.28$-$223210.6 ([IBR2011]\,J1755$-$2232) & 17:55:26.2848 & $-$22:32:10.617 & $-$1.57 & $-$0.07  & 2.0 & 0.107 & 0.001     \\
\scriptsize J180740.68$-$250625.9 ([IBR2011]\,J1807$-$2506) & 18:07:40.6877 & $-$25:06:25.943 & 1.26      & $-$2.64  & 2.0 & 0.059 & 0.001  \\
\scriptsize J175141.34$-$195047.5 ([IBR2011]\,J1751$-$1950) & 17:51:41.3438 & $-$19:50:47.506 & $-$2.43 &      2.62  & 2.0 & 0.012 & 0.001   \\

\hline
\end{tabular}
\end{center} 
}
\end{table*}


\begin{table*}
{\footnotesize
\caption{\textbf{G007.47$+$00.05 -- Parallax \& Proper Motion modeled values.}
Column\,1 reports the LSR velocity of the reference maser spots at ch.\,997 and 1001, respectively; column 2 indicates
the calibration source whose data were used to model the relative proper motion of the masers; column 3 reports the fitted parallax; 
columns 4 and 5 give the fitted proper motions along the east and north directions, respectively. In the last row, we report the distance
measurement inferred from the combined parallax fit. The Galactic proper motion of G007.47$+$00.05 was obtained by the combined fit
value corrected for the internal proper motions of the maser source.}
\begin{center}
\begin{tabular}{ccccc}
\hline \hline
Maser V$_{\rm LSR}$  & Calibration Source & Parallax & $\mu_{\rm x}$  & $\mu_{\rm y}$ \\
(km\,s$^{-1}$)           &                              &   (mas)  & (mas yr$^{-1}$) & (mas yr$^{-1}$)  \\
\hline
$-15.6$ & J175526.28$-$223210.6 & $0.052 \pm 0.007$ & $-2.454 \pm 0.015 $ & $-4.540 \pm 0.303 $  \\
$-16.0$ & J175526.28$-$223210.6 & $0.044 \pm 0.005$ & $-2.511 \pm 0.015 $ & $-4.516 \pm 0.303 $  \\
\multicolumn{5}{l}{\textbf{Combined parallax \& corrected Galactic proper motion:}} \\
\multicolumn{2}{c}{D\,$=$\,20.4$_{-2.2}^{+2.8}$\,kpc} & $0.049 \pm 0.006$ & $ -2.418 \pm  0.017 $ & $ -4.639 \pm  0.303 $  \\
\hline
\end{tabular}
\end{center} 
}
\end{table*}


\begin{table*}
{\footnotesize
\caption{\textbf{Parameters of the 22.2\,GHz H$_2$O maser cloudlets detected towards G007.47$+$00.05.}
Each maser emission center is labeled by decreasing LSR velocity in Column\,1. Columns\,2 and~3 report the LSR
velocity and peak brightness of each cloudlet at the first epoch of detection. Columns\,4 and~5 give the relative
centroid position of each cloudlet, and their uncertainties, in the east and north directions, respectively. The
absolute position of the reference spot, belonging to cloudlet num.\,16, is reported in Table\,S1.
Details for the calculations of the maser cloudlet properties are given in \emph{(31)}. Columns\,6 and~7
give the proper motion components of the cloudlets, with respect to the star, along the east and north directions,
respectively. The relative position of the central star, with respect to cloudlet num.\,16, is also indicated.}
\begin{center}
\begin{tabular}{r r r r r | r r }
\hline \hline
\multicolumn{1}{c}{Feature} & \multicolumn{1}{c}{V$_{\rm LSR}$} & 
\multicolumn{1}{c}{F$_{\rm peak}$} & \multicolumn{1}{c}{$\Delta \rm x$} & \multicolumn{1}{c}{$\Delta \rm y$} & 
\multicolumn{1}{c}{V$_{\rm x}$} & \multicolumn{1}{c}{V$_{\rm y}$} \\ 
\multicolumn{1}{c}{\#} & \multicolumn{1}{c}{(km\,s$^{-1}$)} & 
\multicolumn{1}{c}{(Jy\,beam$^{-1}$)} & \multicolumn{1}{c}{(mas)} & \multicolumn{1}{c}{(mas)} & 
\multicolumn{1}{c}{(km\,s$^{-1}$)} & \multicolumn{1}{c}{(km\,s$^{-1}$)}  \\
\hline

 1   &--5.9      &     0.26   &$-1.39   \pm 0.05 $   &$-15.76  \pm  0.07 $  & ... & ...    \\
 2   &--7.6      &     0.22   &$26.79  \pm  0.09  $  &$-77.76 \pm   0.11 $ & ... & ...    \\
 3   &--10.0    &     0.53   &$26.40  \pm  0.02$    &$-77.90  \pm  0.03 $ & ... & ...     \\
 4   &--10.7    &     0.24   &$5.13  \pm  0.05 $   &$-23.63  \pm  0.09$  & ... & ...    \\ 
 5   &--11.4    &     1.41   & $49.46   \pm 0.01$    &$-51.75   \pm 0.01 $    &$-1.1   \pm  1.4  $  &$-13.4  \pm   1.5$   \\ 
 6   &--12.7    &     0.94   & $49.34   \pm 0.01 $   &$-51.70  \pm  0.01$     &$ 4.5  \pm   1.9 $  & $-11.4 \pm   2.0 $  \\ 
 7   &--13.4    &     0.25   &$-1.81   \pm 0.07 $   &$-9.95   \pm 0.09 $ & ... & ...    \\
 8   &--13.5    &     0.50   &$56.13   \pm 0.03$    &$-45.47  \pm  0.04 $    &$-2.7  \pm   4.3 $   &$-13.7  \pm   5.3 $  \\
 9   &--13.7    &     0.50   &$49.11  \pm 0.02 $  & $-51.57  \pm  0.02 $   &$ 21.6 \pm   10.6 $  & $-19.6  \pm  13.3 $  \\
10  &--14.1    &     0.31   & $0.23   \pm 0.05$    &$-18.35  \pm  0.07$  & ... & ...    \\
11  &--14.9    &     0.27   &$6.60  \pm  0.10 $    &$1.85  \pm  0.12$  & ... & ...    \\  
12  &--15.1    &     0.51   & $ 6.61  \pm  0.03 $   & $1.91  \pm  0.03$  & ... & ...     \\ 
13  &--15.1    &     0.33   & $ 2.24   \pm 0.08 $  &$ -5.21  \pm  0.10$  & ... & ...    \\ 
14  &--15.3   &     0.29   & $-3.97   \pm 0.08 $  &$ -7.55  \pm  0.13 $ & ... & ...    \\
15  &--15.4    &     0.37   &$26.80  \pm  0.07 $  &$ -89.68  \pm  0.12 $ & ... & ...    \\ 
16  &--15.6    &   12.85   & $   0.00\pm0.01$ & $   0.01\pm0.01$ & $  -3.5\pm0.7$  & $  9.8\pm0.8$   \\
17  &--15.7    &     0.82   &$4.42  \pm  0.03$     &$1.65  \pm  0.04 $ & ... & ...    \\
18  & --15.9   &     2.56   & $-4.88 \pm  0.02 $ &$ -1.49  \pm  0.02$  & $ 30.7  \pm   1.6$  & $ 38.0   \pm  1.8 $   \\
19  &--16.0    &    6.08    & $  -9.26\pm0.01$ & $  -6.77\pm0.01$ & $ -13.5\pm0.8$ & $ -6.2\pm0.9$   \\
20  &--16.0    &     0.42   &$ -3.84   \pm 0.06 $  & $-17.79  \pm  0.10 $  & ... & ...    \\
21  &--16.1    &     0.39   &$76.43   \pm 0.09 $   &$-82.41  \pm  0.12$  & ... & ...    \\
22  &--16.4    &     1.32   &$39.19   \pm 0.02 $   &$-43.06   \pm 0.02 $     &$3.2  \pm   2.1 $   & $-8.7   \pm  2.2 $   \\
23  &--16.4    &     0.57   &$113.44   \pm 0.02$   &$ -100.42  \pm  0.02 $  & ... & ...      \\
24  &--16.6    &     0.40   &$-5.17  \pm  0.04$   &$ -2.05  \pm  0.06$  & ... & ...    \\
25  &--16.8    &     0.37   &$4.26  \pm  0.03 $   &$-8.39  \pm  0.04$  & ... & ...    \\
26  &--17.1    &     0.74   &$-5.29   \pm 0.02 $   &$-2.14  \pm  0.03 $   &$-36.0  \pm   7.9 $  &  $ 9.0  \pm  10.2 $  \\
27  &--17.2    &     0.29   &$48.09  \pm  0.04$    &$-36.45  \pm  0.07 $  & ... & ...      \\
28  &--17.2    &     0.20   &$43.74   \pm 0.14 $   &$-43.09  \pm  0.15 $  & ... & ...      \\
29  &--18.2    &     0.74   &$43.85  \pm  0.02  $  &$-43.45  \pm  0.02  $   &$22.2   \pm  2.8  $   &$12.1  \pm   3.1$   \\
30  &--18.2    &     0.56   & $38.14   \pm 0.03 $   &$-43.07  \pm  0.03$  & ... & ...      \\
31  &--18.4    &     1.93   & $-1.92   \pm 0.01 $  & $-11.63  \pm  0.01  $ & $-33.7 \pm 1.1 $  & $-11.3  \pm   1.2 $  \\
32  &--19.3    &     1.02   &$ -1.97   \pm 0.01  $ &$ -11.70  \pm  0.01 $   & $-9.2  \pm   1.8$    & $ 0.5 \pm   1.9$   \\
33  &--21.9    &     0.37   &$-2.05   \pm 0.03 $   &$-11.68  \pm  0.05 $  & ... & ...    \\
34  &--22.6    &     0.27   & $43.12  \pm  0.05$    &$-29.06  \pm  0.08 $ & ... & ...    \\
35  &--22.5    &     0.53   & $43.20  \pm  0.03 $   &$-29.02  \pm  0.03$  & ... & ...    \\
36  &--24.9    &     0.57   &$46.48   \pm 0.02  $  &$-41.52   \pm 0.02$  & ... & ...    \\

& & & & & & \\

Star & & &  $21.99\pm0.01$ & $-26.70\pm0.01$ & & \\

\hline
\end{tabular}
\end{center} 
}
\end{table*}

\end{document}